\documentclass[aps,amssymb,showpacs]{revtex4-1}
\usepackage{graphicx}
\usepackage{color}

\begin{document}
\def\be{\begin{equation}}
\def\ee#1{\label{#1}\end{equation}}

\newcommand{\ben}{\begin{eqnarray}}
 \newcommand{\een}{\end{eqnarray}}
  \def\no{\nonumber}
\def\lb{\label}

\title{Isotropization in Bianchi type-I cosmological model \\ with fermions and bosons interacting via Yukawa potential}
\author{M. O.  Ribas}\email{gravitam@yahoo.com}
\affiliation{Departamento de F\'{\i}sica, Universidade Tecnol\'ogica Federal do
Paran\'a, Curitiba, Brazil}
\author{L. L. Samojeden} \email{samojed@fisica.ufpr.br}
\affiliation{Departamento de F\'{\i}sica, Universidade Federal do
Paran\'a,  Curitiba, Brazil}
\author{F. P. Devecchi}\email{devecchi@fisica.ufpr.br}
\affiliation{Departamento de F\'{\i}sica, Universidade Federal do
Paran\'a,  Curitiba, Brazil}
\author{G. M. Kremer}\email{kremer@fisica.ufpr.br}
\affiliation{Departamento de F\'{\i}sica, Universidade Federal do
Paran\'a,  Curitiba, Brazil}

\begin{abstract}
In this work  we  investigate a model for the early Universe in a Bianchi type-I metric, where the sources of the gravitational field  are a fermionic and a bosonic field, interacting through a Yukawa potential, following the standard model of elementary particles. It is shown  that the fermionic field has a negative pressure, while the boson has a small positive pressure. The fermionic field is  the responsible for an accelerated regime at early times, but since the total  pressure tends to zero for large times, a transition to a decelerated regime occurs. Here the Yukawa potential answers  for the duration of the accelerated regime, since by decreasing  the value of its coupling constant the transition accelerated-decelerated occurs in later times.   The isotropization which occurs for late times is due to the presence of the fermionic field as one of the sources of the gravitational field.

\end{abstract}
 \maketitle

\section{Introduction}

The majority of cosmological models which describe the early, intermediate and present  periods of the Universe are based
 on  observational data which suggest that the Universe is homogeneous and
isotropic at large  scales. The geometry adopted to describe these scenarios at large scales is the
Friedmann-Lama\^itre-Robertson-Walker (FLRW) metric. However, high precision measurements   conducted by the Cosmic
Background Explorer (COBE) \cite{1} and by the Wilkinson Microwave Anisotropy Probe (WMAPE)
\cite{2} show small fluctuations in the temperature amplitude in different directions, which indicate a small anisotropy
of the young Universe.  Hence, the analysis of an early anisotropic Universe and the mechanisms which lead to a
posterior isotropization is a topic of theoretical interest.

Recently several cosmological models were proposed in the literature, where fermionic fields acting as gravitational
sources lead to accelerated regimes. Among others we quote the works
\cite{s1,s2,ar,s3,r1,s4,r2,rd,sm,r3,r4,s5}. In some of these works it is also discussed  the role of a fermionic field in
 the isotropization which occurs in a Bianchi type-I metric (see e.g. \cite{s1,s2,s3,s4,s5}).

The aim of this work is to describe the evolution of a young accelerated Universe from an anisotropic scenario to an almost complete isotropization observed in the matter dominated Universe. To that end we consider as the sources of the gravitational field a fermionic and  a bosonic field interacting via a Yukawa potential - following elementary particle theory\cite{Ryder} -  and a Bianchi type-I metric -- which is spatially homogeneous and represents an anisotropic generalization of the FLRW metric.

The role of fermionic and bosonic fields interacting via a Yukawa potential in a FLRW metric was discussed previously in the work \cite{r3}, where it was shown that this cosmological model could describe the accelerated-decelerated transition of the early Universe, with the Yukawa coupling controlling the duration of the accelerated period. Here we are interested in studing the same model within the framework of a Bianchi type-I metric. Among other results it is shown that the fermionic field is the responsible for the initial accelerated regime, since its pressure is negative and large in modulo. This overpowers the bosonic pressure, which  is positive. For large times the sum of the fermionic and bosonic pressures tend to zero and the Universe enters in a decelerated period, which can be identified with a matter (dust) dominated period. The  coupling constant of the Yukawa potential answers also  for the duration of the accelerated regime, due to the fact that the  increase of the coupling constant implies that the accelerated-decelerated transition occurs at earlier times. It is important to call attention that in the absence of the Yukawa potential only an accelerated regime takes place.   Furthermore, as was previously investigated by Saha \cite{s5},   the isotropization is due to the presence of the fermionic field as the source of the gravitational field.

The manuscript is
structured as follows: in section II we present the cosmological model where a fermionic and bosonic field interact through a Yukawa potential and derive the Klein-Gordon, Dirac and Einstein field equations. These equation are written in a Bianchi type-I metric in section III. In Section IV we derive the cosmological solutions through a numerical analysis of the field equation and discuss the role of the fermionic and bosonic fields, the Yukawa interaction and the isotrpization conditions. In the last section the main conclusions of the work are stated. The metric signature used is  $(+,-,-,-)$ and units
have been chosen so that $8\pi G=c=\hbar=1$.

\section{The model}

In this work we are interested in investigating a cosmological model in a Bianchi type-I space-time, where the sources of the gravitational field are a fermion and a boson which are interacting
via a Yukawa potential, following the standard model of elementary particles\cite{Ryder}.

 We start by focusing on the fermionic sector, where the tetrad formalism  solves  the spinor representation problem
in general relativity \cite{Weinberg,Wald,Ryder,Bir}.
By invoking the equivalence principle -- where to each space-time point we can associate a local inertial frame -- the connection of the local inertial frame with the space-time structure is  established by introducing the tetrad field $e^\mu_a=\partial x^\mu/\partial \xi^a$. Here the coordinates $x^\mu$ are related to the differentiable manifold which represents the the space-time and $\xi^a$ to the tangent space. The Latin indexes correspond to the local Lorentz frame, while the Greek ones to the general frame. The relationship between the metric tensor $g_{\mu\nu}$ and the tetrad $e^a_\mu$ is given by $g_{\mu\nu}=e_\mu^a e_\nu^b\eta_{ab}$, where $\eta_{ab}$ is the Minkowski metric tensor.

The general covariance principle\cite{Weinberg}
imposes that the Dirac-Pauli matrices $\gamma^a$ give place to
their generalized versions $\Gamma ^{\mu}=e^\mu_a\gamma^a$, where
the new matrices satisfy the  extended
Clifford algebra,  $ \{\Gamma^\mu,\Gamma^\nu\}=2g^{\mu\nu}.$ Furthermore, the ordinary derivatives are replaced  by their covariant counterparts
\be
\partial_\mu\psi\rightarrow D_\mu\psi= \partial_\mu\psi-\Omega_\mu\psi,\quad
\partial_\mu\overline\psi\rightarrow
D_\mu\overline\psi=\partial_\mu\overline\psi+\overline\psi\Omega_\mu.
\ee{1}
Here $\Omega_\mu$ denotes the spin connection
\ben\lb{2}
\Omega_\mu=-\frac{1}{4}g_{\rho\sigma}\left[\Gamma^\rho_{\mu\delta}
-e_b^\rho\left(\partial_\mu e_\delta^b\right)\right]\Gamma^\delta\Gamma^\sigma,
\een
and $\Gamma^\rho_{\mu\delta}$ are Christoffel symbols.

The next step  is to couple the fermionic field to gravity by constructing
a generally covariant Dirac Lagrangian density \cite{Bir} for a massless spinor field
\ben
L_f=\frac{i}{2}\left[ \overline\psi\,\Gamma^\mu
D_\mu\psi-(D_\mu\overline\psi)\Gamma^\mu\psi\right]-V\left(\overline\psi\psi\right).
\label{3}
\een
Here $\psi$ and $\overline\psi=\psi^\dag\gamma^0$ are the spinor field and its adjoint, respectively, while $V(\overline\psi\psi)$ is  the  fermionic self-interaction potential, which will be assumed as a function of the bilinear $\overline\psi\psi$.

The Lagrangian density of a real massive scalar field $\phi$ is given by \cite{Ryder}
\ben\lb{4}
L_b=\frac12\partial_\mu\phi\partial^\mu\phi-\frac12m\phi^2,
\een
where $m$ denotes the bosonic mass.

Here we follow \cite{r3} and assume that the fermionic and bosonic fields are interacting through a Yukawa potential. In this case the Lagrangian density corresponding to this interaction reads
\ben\lb{5}
L_Y=-\lambda\overline\psi\phi\psi.
\een
Above, $\lambda$ is the coupling constant.

From the  total action
\ben S=\int \sqrt{-g}\left\{\frac12 R+L_f+L_b+L_Y\right\}d^4 x,\label{6}
\een
where $R$ denotes the scalar curvature, we obtain the general field equations for this model.
First from the variation of the action (\ref{6}) with respect to the spinor field and its adjoint we get Dirac's equations coupled to the Yukawa potential:
\ben\lb{7a}
(D_\mu\overline\psi)\Gamma^\mu-i\lambda\overline\psi\phi-i\frac{dV}{d\psi}=0,\\\lb{7b}
\Gamma^\mu D_\mu\psi+i\lambda\psi\phi+i\frac{dV}{d\overline\psi}=0.
\een
Next the Klein-Gordon equation coupled with the Yukawa potential follows from the variation of the action (\ref{6}) with respect to the scalar field:
\ben\lb{8}
\nabla_\mu\nabla^\mu\phi+m^2\phi+\lambda\overline\psi\psi=0.
\een

 Einstein's field equations are obtained from the variation of the action (\ref{6}) with respect to the tetrad, yielding,
 \ben\lb{9}
 R_{\mu\nu}-\frac{1}{2}g_{\mu\nu}R=-T_{\mu\nu}.
 \een
 Here $T_{\mu\nu}$ is the energy-momentum tensor of the gravitational sources, which is given by
\ben\no
T^{\mu\nu}=\frac{i}{4}\left[\overline\psi\Gamma^\mu D^\nu\psi+\overline\psi\Gamma^\nu D^\mu\psi-D^\nu\overline\psi\Gamma^\mu\psi-D^\mu\overline\psi\Gamma^\nu\psi\right]
\\\lb{10}
+\nabla^\mu\phi\nabla^\nu\phi-g^{\mu\nu}\left(L_f+L_b+L_Y\right).\qquad\qquad
\een

\section{Field equations in Bianchi type-I metric}

Let us study the dynamics of a homogeneous but anisotropic primordial Universe. To that end we shall use a Bianchi type-I metric described by the line element
\ben\lb{11}
ds^2=dt^2-a(t)^2dx^2-b(t)^2dy^2-c(t)^2dz^2.
\een
The homogeneity is granted by the independence of the scale factors $a(t)$, $b(t)$ e $c(t)$ on the spatial coordinates, while the anisotropy follows from the three different scale factors in the spatial directions.

For the  Bianchi type-I metric (\ref{11}) the components of the tetrad, spin connection and Dirac-Pauli matrices read
\ben\no
e^\mu_0=\delta^\mu_0, \quad e^\mu_1=\frac{1}{a(t)}\delta^\mu_1,\quad e^\mu_2=\frac{1}{b(t)}\delta^\mu_2, \\\no
e^\mu_3=\frac{1}{c(t)}\delta^\mu_3,\quad\Omega_0=0,\quad \Omega_1=\frac{1}{2}\dot a(t)\gamma^1\gamma^0, \\\no
\Omega_2=\frac{1}{2}\dot b(t)\gamma^2\gamma^0, \quad \Omega_3=\frac{1}{2}\dot c(t)\gamma^3\gamma^0,\quad\Gamma^0=\gamma^0,\\\no
  \Gamma^1=\frac{1}{a(t)}\gamma^1, \quad \Gamma^2=\frac{1}{b(t)}\gamma^2, \quad \Gamma^3=\frac{1}{c(t)}\gamma^3.
\een

In the Bianchi type-I metric the  Dirac (\ref{7a}) and (\ref{7b}) and Klein-Gordon (\ref{8}) equations become
\ben\lb{12a}
\dot\psi+\frac{1}{2}\left(\frac{\dot a }{a}+\frac{\dot b}{b}+\frac{\dot c}{c}\right)\psi
+iV'\gamma^0\psi+i\lambda\phi\gamma^0\psi=0,
\\\lb{12b}
\dot{\overline\psi}+\frac{1}{2}\left(\frac{\dot a }{a}+\frac{\dot b}{b}+\frac{\dot c}{c}\right)\overline\psi-iV'\overline\psi\gamma^0-i\lambda\phi\overline\psi\gamma^0=0,
\\\lb{12c}
\ddot\phi+\left(\frac{\dot a }{a}+\frac{\dot b}{b}+\frac{\dot c}{c}\right)\dot\phi+m^2\phi+\lambda\overline\psi\psi=0.
\een
Above $V'$ denotes the derivative of the self-interaction potential $V(\overline\psi\psi)$ with respect to the bilinear, i.e., $V'={dV}/{d(\overline\psi\psi)}$.

The diagonal elements of Einstein's field equations (\ref{9}) in the Bianchi type-I metric can be written as
\ben\lb{13a}
\frac{\dot a\dot b}{ab}+\frac{\dot a \dot c}{ac}+\frac{\dot c\dot b}{bc}=\rho,
\\\lb{13b}
\frac{2\ddot c}{c}+\left(\rho-\frac{2\dot a\dot b}{ab}\right)=-p,
\\\lb{13c}
\frac{2\ddot b}{b}+\left(\rho-\frac{2\dot a\dot c}{ac}\right)=-p,
\\\lb{13d}
\frac{2\ddot a}{a}+\left(\rho-\frac{2\dot b\dot c}{bc}\right)=-p.
\een
We call attention to the fact that the diagonal components of the energy-momentum tensor $T_1^1=T_2^2=T_3^3$ so that we may write  $T^\mu_\nu=(\rho,-p,-p,-p)$, where $\rho$ and $p$ are the energy density and pressure of the sources of the gravitational field, respectively.
The expressions for the energy density and pressure are given by
\ben\lb{14a}
\rho=\frac{1}{2}\dot\phi^2+\frac{1}{2}m^2\phi^2+V(\overline\psi\psi)+\lambda\overline\psi\psi\phi,\\\lb{14b}
p=\frac{1}{2}\dot\phi^2-\frac{1}{2}m^2\phi^2+V'(\overline\psi\psi)\,\overline\psi\psi-V(\overline\psi\psi).
\een

From the above equations we can identify (\ref{13a}) as the Friedmann equation, (\ref{13b}) -- (\ref{13d}) as the acceleration equations.  

We may interpret the total energy density of the sources of the gravitational field as a sum of three contributions $\rho=\rho_b+\rho_f+\rho_Y$ which are related with the bosonic, fermionic fields and Yukawa potential, respectively. Their expressions read
\be
\rho_b=\frac{1}{2}\dot\phi^2+\frac{1}{2}m^2\phi^2,\qquad  \rho_f=V(\overline\psi\psi),\qquad
\rho_Y=\lambda\overline\psi\psi\phi.
\ee{15}
In the same way we may interpret the total pressure of the sources of the gravitational field as a sum of a bosonic and a fermionic pressures, $p=p_b+p_f$, whose expressions are given by
\be
p_b=\frac{1}{2}\dot\phi^2-\frac{1}{2}m^2\phi^2,\qquad
p_f=V'(\overline\psi\psi)\,\overline\psi\psi-V(\overline\psi\psi).
\ee{16}
Note that the fermionic energy density and pressure are due to its self-interaction potential.

As was pointed by Saha \cite{s5}, while the non-diagonal components of the left-hand side Einstein's field equations (\ref{9}) vanish, the components of the energy-momentum tensor on the right-hand side are not zero. Indeed, the components $T_{12}$, $T_{13}$ and $T_{23}$ read
\ben\lb{17a}
\frac{b}{a}\left(\frac{\dot a}{a}-\frac{\dot b}{b}\right)\overline\psi\gamma^1\gamma^2\gamma^0\psi=0,
\\\lb{17b}
\frac{c}{a}\left(\frac{\dot c}{c}-\frac{\dot a}{a}\right)\overline\psi\gamma^3\gamma^1\gamma^0\psi=0,
\\\lb{17c}
\frac{c}{b}\left(\frac{\dot b}{b}-\frac{\dot c}{c}\right)\overline\psi\gamma^2\gamma^3\gamma^0\psi=0.
\een
 If we impose that the terms inside the braces of the above equations vanish, we get that the anisotropy rules out, since $a(t)=b(t)=c(t)$. This case was previously analyzed in the work \cite{r3}. Hence, the only condition we can impose is that
\ben\lb{18}
\overline\psi\gamma^1\gamma^2\gamma^0\psi=0,\quad
\overline\psi\gamma^3\gamma^1\gamma^0\psi=0,\quad
\overline\psi\gamma^2\gamma^3\gamma^0\psi=0,\qquad
\een
which are algebraic constraints on the components of the spinor field. The algebraic constraints (\ref{18}) in terms of the components of the spinor field can be written as
\ben\no
\psi_1^*\psi_2-\psi_2^*\psi_1+\psi_3^*\psi_4-\psi_4^*\psi_3=0,\\\no
\psi_1^*\psi_2+\psi_2^*\psi_1+\psi_3^*\psi_4+\psi_4^*\psi_3=0,\\\no
\psi_1^*\psi_1-\psi_2^*\psi_2-\psi_3^*\psi_3+\psi_4^*\psi_4=0.
\een
However, for the solution of the field equations it will not necessary to use the above constraints, since the only quantity that appear in the field equations is the bilinear $\overline\psi\psi$ which is an explicit function of the cosmic scale factors. Indeed, the sum of the right multiplication of (\ref{12a}) by $\overline\psi$ together with the left multiplication of (\ref{12b}) by $\psi$ leads to
\ben
d\ln\left(\overline\psi\psi\right)+d\ln\left(abc\right)=0,
\een
whose solution is
\ben
\left(\overline\psi\psi\right)(t)=\frac{\mathcal{C}_1}{a(t)b(t)c(t)}.
\een
Here $\mathcal{C}_1$ is an integration constant.

Once the bilinear is known as a function of the cosmic scale factors, we need to determine the time evolution of the cosmic scale factors and of the scalar field. To this end we shall use the Klein-Gordon (\ref{12c}) and the acceleration equations (\ref{13b}), (\ref{13c}) and (\ref{13d}).
This system of equations  constitute a highly non-linear coupled system of differential
 equations and we proceed to solve it numerically in the next section.

 \section{Cosmological solutions}

In order to solve the coupled system of equations (\ref{12c}), (\ref{13b}), (\ref{13c}) and (\ref{13d}) we have to
specify the self-interaction potential of the spinor field $V(\overline\psi\psi)$ and the initial conditions for the
three cosmic scale factors, for the scalar field and for their first derivatives with respect to time. We follow \cite{r3} and suppose that the self-interaction potential of the spinor field is given by
$V(\overline\psi\psi)={\mathcal{C}_2}(\overline\psi\psi)^n,$
where ${\mathcal{C}_2}$ and $n$ are constants.
In the numerical simulations the adopted values for the cosmic scale factors were:
\ben\no
a(0)=9.185,\quad  b(0)=9.12, \quad c(0)=9.267,\\\no \dot a(0)=0.1021, \quad\dot b(0)=0.103,\quad \dot c(0)=0.101,
\een
where a small anisotropy in their values and in their time derivatives were taken into account.
For the scalar field the adopted value was $\phi(0)=0.5$, while the value of $\dot\phi(0)$ was determined from the Friedmann equation, since according to (\ref{13a}) we have
\ben\no
\dot\phi(0)^2=2\frac{\dot a(0)\dot b(0)}{a(0)b(0)}+\frac{2\dot a(0)\dot c(0)}{a(0)c(0)}+\frac{2\dot b(0)\dot c(0)}{b(0)c(0)}\\-\frac{2\lambda \mathcal{C}\phi(0)}{a(0)b(0)c(0)}-m^2\phi(0)^2-\frac{2\mathcal{C}}{[a(0)b(0)c(0)]^{n}},
\een
where $\mathcal{C}=\mathcal{C}_1{\mathcal{C}_2}$ is a constant.

\begin{figure}
\vskip1cm
\includegraphics[width=10cm]{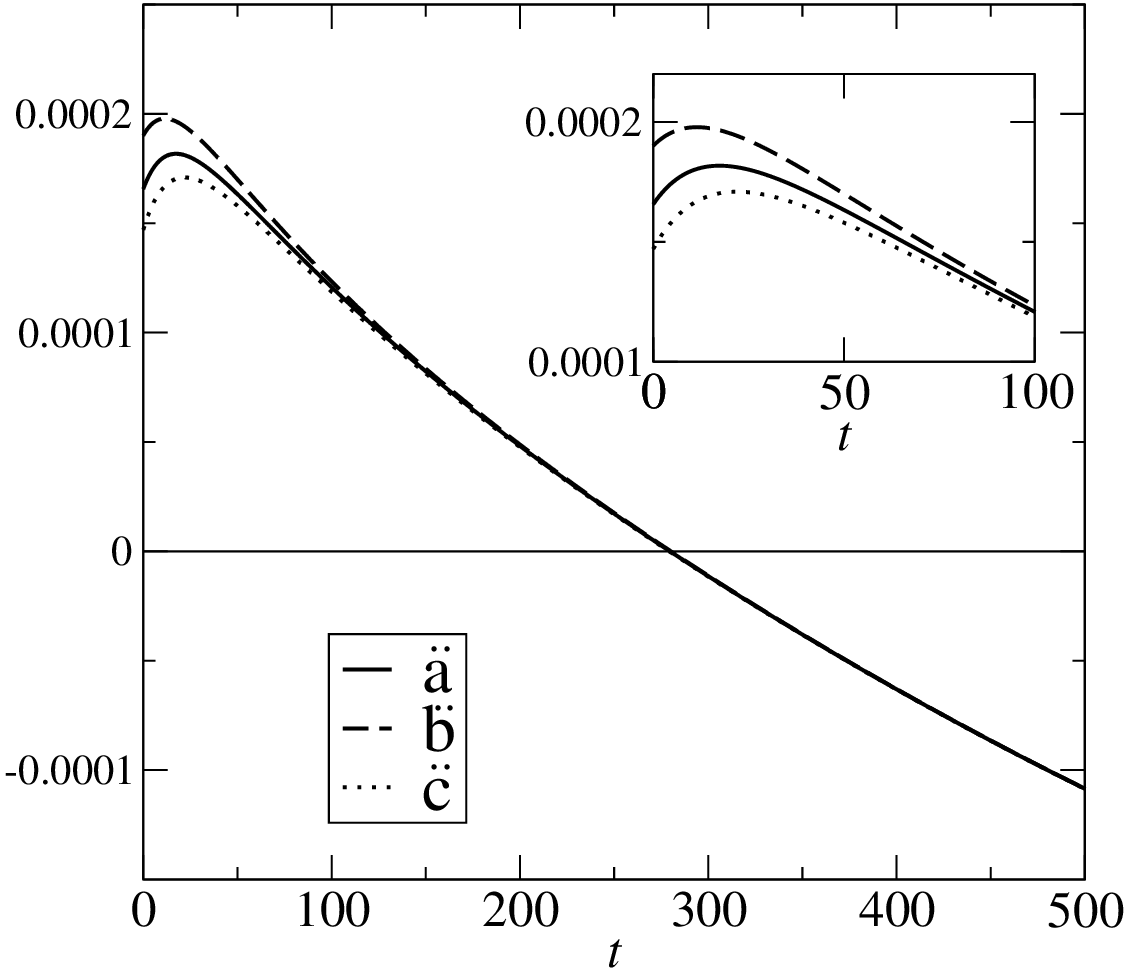}
\caption{Accelerations as functions of time. }
\end{figure}

Apart from the initial conditions we have also to specify: the coupling constant $\lambda$ of the Yukawa potential,
the exponent $n$ of the fermionic self-interaction potential, the constant $\mathcal {C}$ and the bosonic mass $m$.
The adopted values were:
\be
\lambda=5\times10^{-3},\qquad n=\frac12,\qquad \mathcal{C}=10^{-2},\qquad m=10^{-4}.
\ee{c}

\begin{figure}
\vskip1cm
\includegraphics[width=10cm]{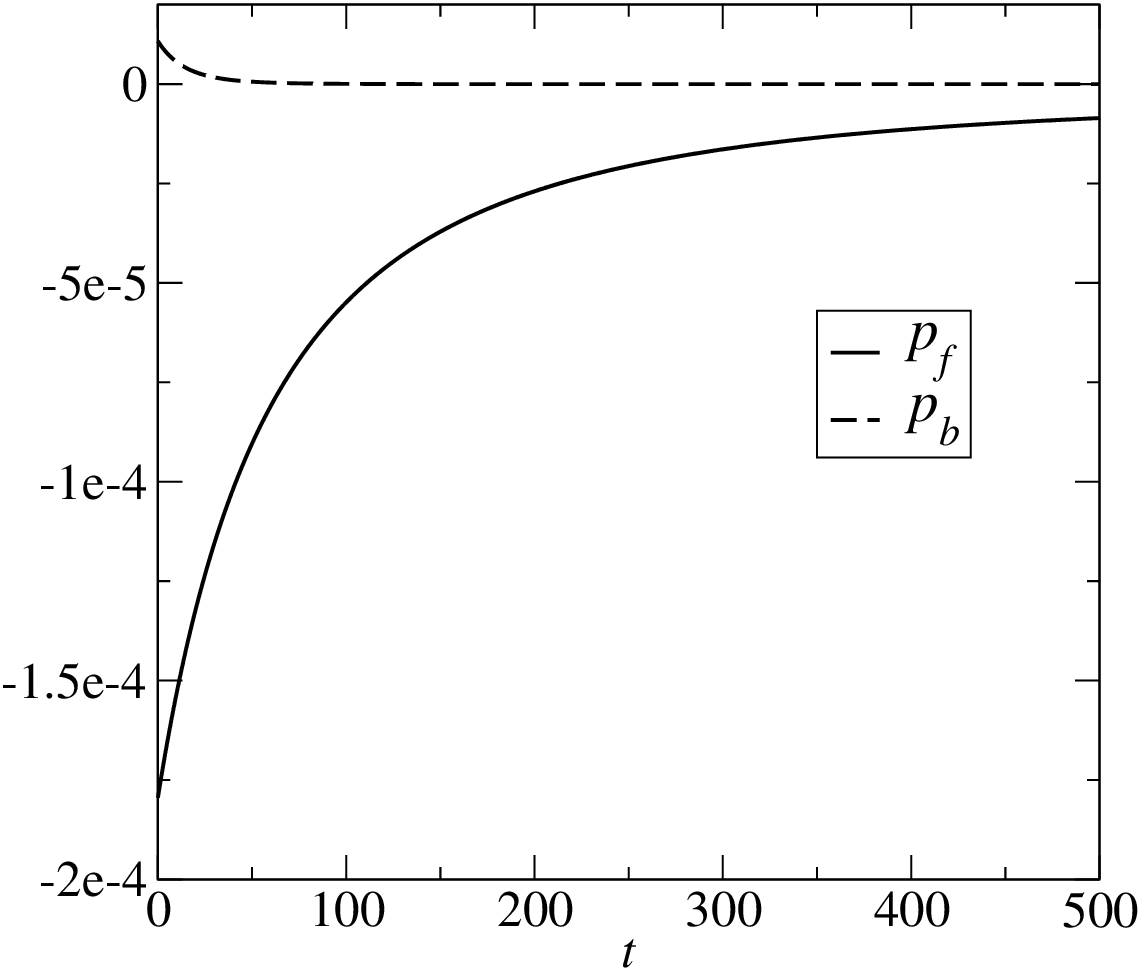}
\caption{Bosonic and fermionic pressures as functions of time.}
\end{figure}

In Fig. 1 the behavior of the accelerations in the three spatial directions are plotted as functions of time.
 We infer from this figure that the initial anisotropy of the accelerations fades away with time. Furthermore, a transition of an accelerated regime to a decelerated one occurs. This fact can be understood by the analysis of Fig. 2 were it is shown the time evolution of the bosonic and fermionic pressures. The bosonic pressure is positive but has a smaller module in comparison with the fermionic one. The fermionic pressure is negative and responsible for the positive acceleration, but it tends to zero with  time behaving as  dust matter,  which corresponds to a decelerated period. Here we call attention to the fact that the role of the Yukawa potential is important in the accelerated-decelerated transition, since if we decrease the value of the coupling constant $\lambda$ this transition occurs in later times. Furthermore, without the Yukawa potential there is no transition and there exists only an accelerated expansion of the Universe.

\begin{figure}
\vskip1cm
\includegraphics[width=10cm]{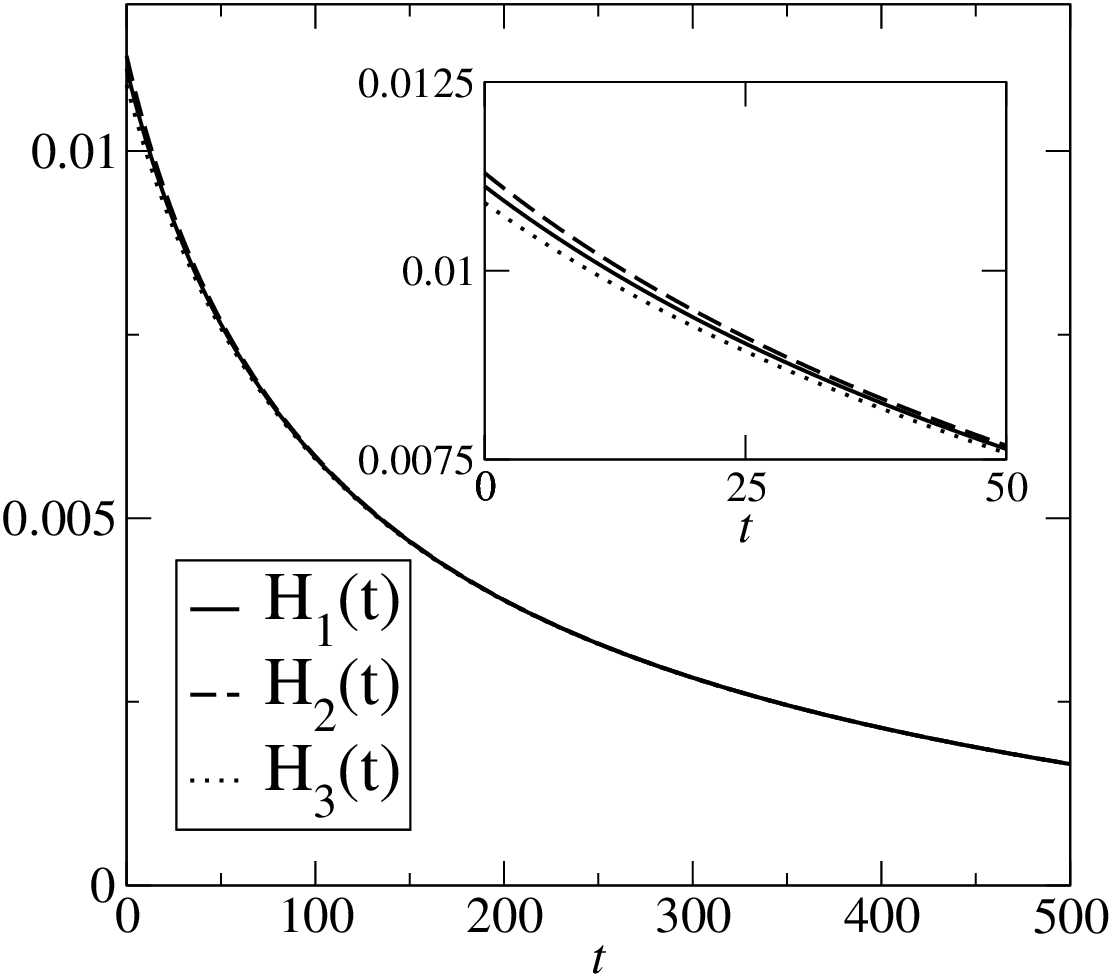}
\caption{Hubble parameters as functions of time.}
\end{figure}

In Fig. 3 the three Hubble parameters
\ben\no
H_1=\frac{\dot a(t)}{a(t)},\qquad H_2=\frac{\dot b(t)}{b(t)},\qquad H_3=\frac{\dot c(t)}{c(t)},
\een
are plotted as functions of time. We note from this figure that the anisotropy of those parameters  also fades away with time.

\begin{figure}
\vskip1cm
\includegraphics[width=7cm]{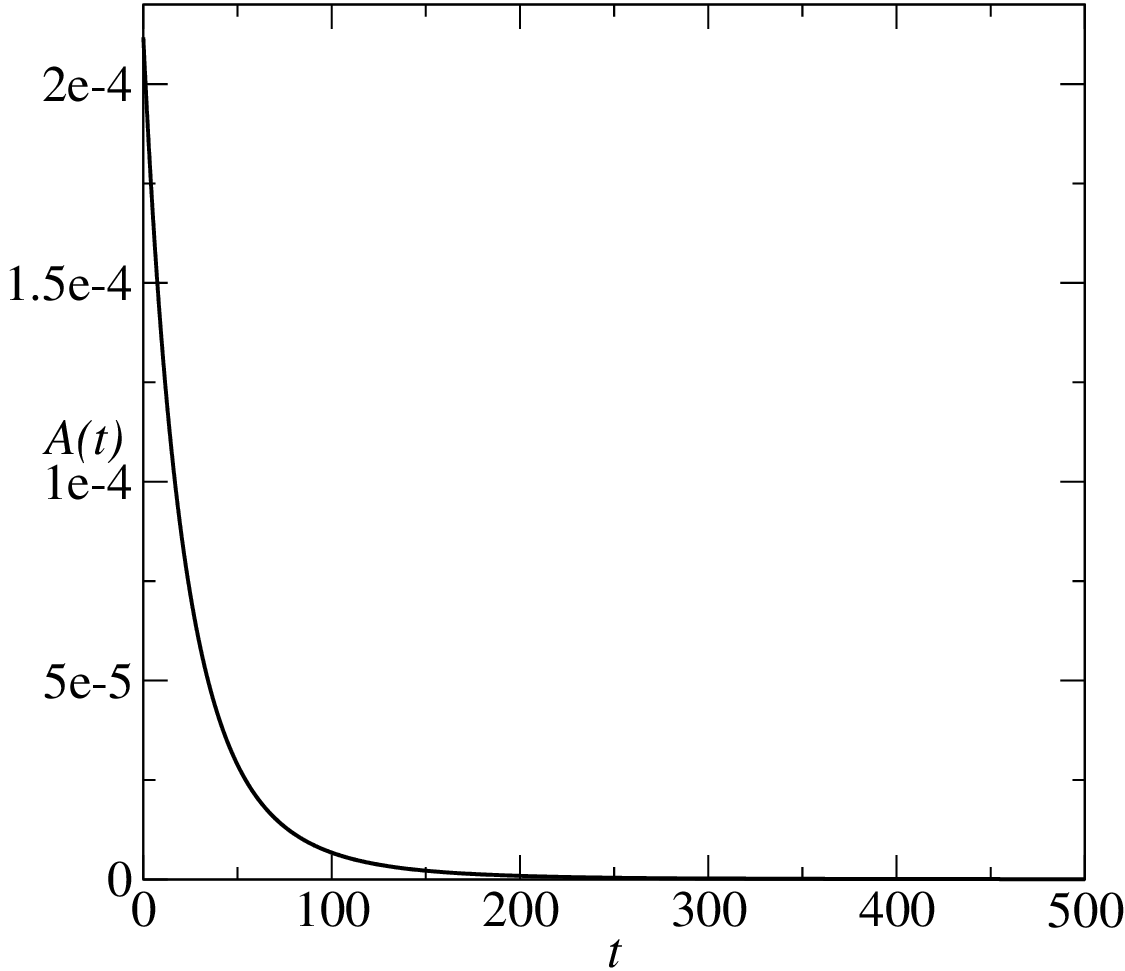}
\caption{Isotropization criterium (\ref{is1}) as function of time.  }
\end{figure}

\begin{figure}
\vskip1cm
\includegraphics[width=7cm]{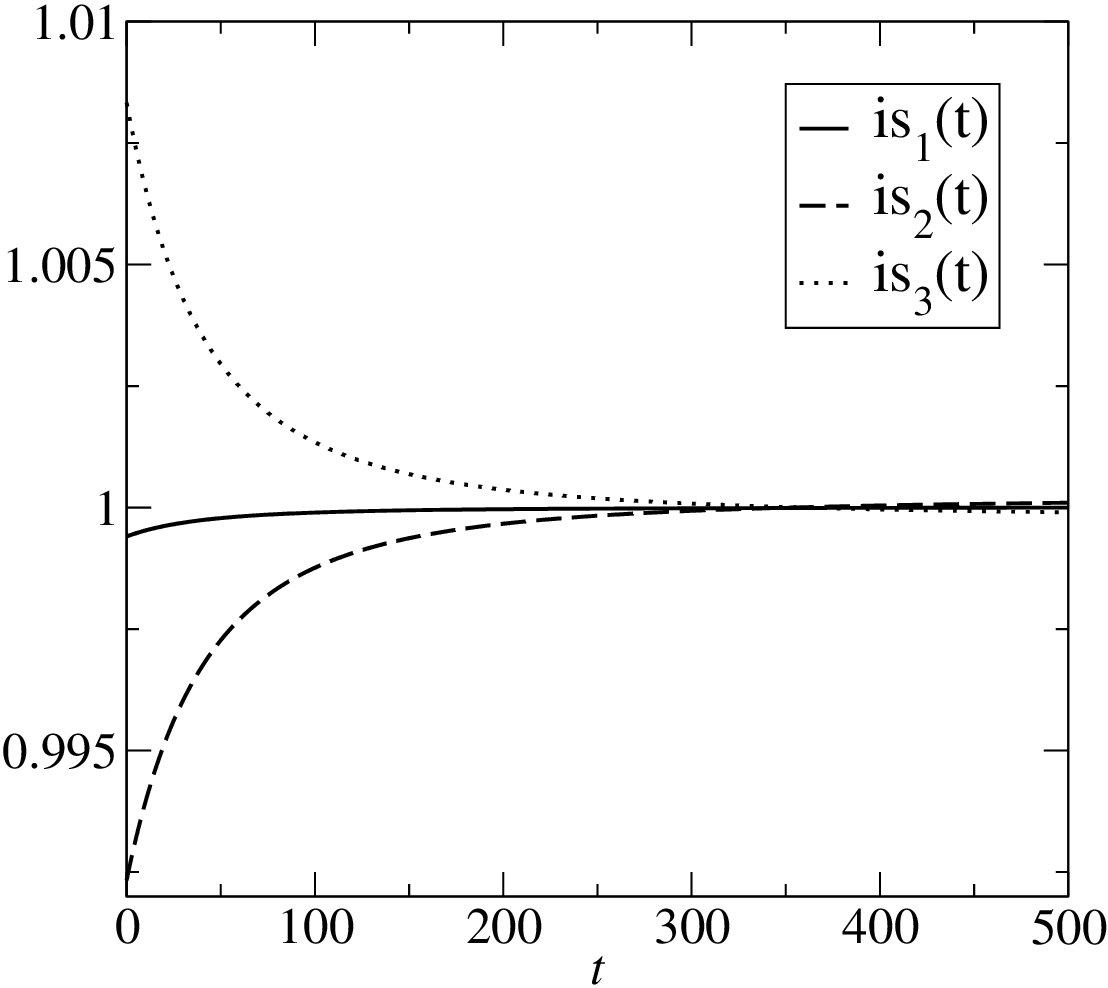}
\caption{Isotropization criteria (\ref{s}) as function of time.}
\end{figure}

Although the figures for the accelerations and Hubble parameters show a tendency to isotropy with time we use two more criteria to confirm that this happens for this model. The first one, proposed by Jacob \cite{JC}, requires that
\ben\lb{is1}
A(t)=\frac{1}{3}\sum_{i=1}^{3}\frac{H_i^2}{H^2}-1\rightarrow0,
\een
tends to zero for large values of time. Here
\ben\no
H=\frac{\dot\tau}{\tau},\qquad\hbox{where}\qquad \tau=\left[a(t)b(t)c(t)\right]^\frac13.
\een
We observe from Fig. 4 that this criterium is verified by this model.
The next criterium is due to Saha \cite{s5}, which imposes that the ratios
\ben\lb{s}
{\rm is}_1(t)=\frac{a(t)}{\tau},\quad {\rm is}_2(t)=\frac{b(t)}{\tau},\quad {\rm is}_3(t)=\frac{c(t)}{\tau},
\een
tend to constant values for large values of time. This is also verified by the present model if one observes the behavior of these ratios as functions of time in Fig. 5.

\section{Conclusions}

In this work we have investigated the dynamics of a fermionic field interacting with a massive bosonic field via a Yukawa potential, as proposed in the standard model, in a homogeneous anisotropic Universe described by the Bianchi type-I metric. The fermionic field promotes the acceleration of the primordial Universe, since it  has a negative pressure whose absolute value is larger than the positive pressure of the bosonic field. Both pressures  tend to zero for large values of time so that the Universe enters  into a matter dominated decelerated period. Without the presence of the bosonic field the cosmological solution is that of de Sitter. The results show that the initial spatial anisotropy of the Universe dilutes with time due to the presence of the Dirac field. Furthermore,  the Yukawa potential is essential in order to have a transition from an accelerated to a decelerated period, without it only an accelerated period is possible.


\end{document}